# Dynamic PET Image Prediction Using a Network Combining Reversible and Irreversible Modules

Jie Sun[a], Junyan Zhang[a], Qian Xia[b], Chuanfu Sun[a],Yumei Chen[b], Yunjie Yang[c]
Huafeng Liu[d], Wentao Zhu[d*], Qiegen Liu[a,*]

[a] school of Information Engineering, Nanchang University, China.
[b] Institute of Molecular Medicine, Renji Hospital, Shanghai Jiao Tong University School of Medicine, China.
[c] Institute for Imaging, Data and Communications, School of Engineering, The University of Edinburgh, Edinburgh, U.K.
[d] College of Biomedical Engineering and Instrument Science, Zhejiang University, China.

**Abstract:**

Dynamic positron emission tomography (PET) images can reveal the distribution of tracers in the organism and the dynamic processes involved in biochemical reactions, and it is widely used in clinical practice. Dynamic PET imaging is useful in analyzing the kinetics and metabolic processes of radiotracers. Prolonged scan times can cause discomfort for both patients and medical personnel. This study proposes a frame prediction method for dynamic PET imaging, reducing PET scanning time by applying a multi-module deep learning framework composed of reversible and irreversible modules. The network can predict kinetic parameter images based on the earlier frames of dynamic PET images, and then generate complete dynamic PET images. In the experiments with simulated data and the generalization experiments with clinical data, the dynamic PET images predicted by our network have higher SSIM and PSNR and lower MSE than its counterparts. The generalization performance of this network

*Corresponding author : Wentao Zhu, Qiegen Liu.
*Email address:* wentao.zhu@zhejianglab.com (Wentao . Zhu), liuqiegen@ncu.edu.cn (Qiegen. Liu).
Jie Sun and Junyan Zhang are co-first authors
416100220068@email.ncu.edu.cn (Jie Sun), zhangjunyan@email.ncu.edu.cn(Junyan Zhang), xiaqian@renji.com (Qian Xia), 416100220243@email.ncu.edu.cn(Chuanfu Sun), chenyumei@renji.com(Yumei Chen), y.yang@ed.ac.uk(Yunjie Yang), liuhf@zju.edu.cn(Huafeng Liu) in clinical data experiments indicates that the proposed method has potential in the application of dynamic PET.

## 1. Introduction

Positron emission tomography (PET) is a powerful nuclear imaging technique used to assess the metabolic function of living organisms. It plays a crucial role in clinical settings for detection, diagnosis, staging, treatment planning, and disease monitoring [1-2]. PET has proven to be effective in assisting with tumor diagnosis, cancer detection [3] and neurological diseases [4]. Due to its high sensitivity and specificity, especially in lesion detection, PET has become an invaluable tool in clinical practice for disease diagnosis and treatment [5-6].

Compared to traditional static PET imaging techniques, dynamic PET imaging has greater clinical utility in tracking disease therapy and tumor diagnosis [7-8]. In dynamic PET studies, mathematical models are widely used to describe the relationship between the measured temporal data and the kinetic physiological parameters that determine the uptake of a radiotracer and its clearance [9]. In this context, compartment models are the most commonly used type of models. By introducing compartment models, dynamic PET data can be transformed into physiological parameter data that reflects the functional status of tissues. Compared to standard care static activity and weight-normalized PET images referred to as standardized uptake value (SUV) images, parametric images exhibit superior performance in lesion detection and characterization [10-11]. The Logan plot, as a type of parametric image, can quantitatively elucidate physiological and biochemical processes [12]. In addition to the Logan plot, the Patlak graphical method is widely used in dynamic PET for modeling the kinetics of tracer parameters imaging [13]. In compartmental models, the exchange of tracer between compartments is simulated using first-order ordinary differential equations, where the coefficients are referred to as kinetic parameters. Obtaining the kinetic parameters of compartmental models is of significant importance. In certain scenarios, a homogeneous tissue region, such as the myocardium or even the entire striatum in brain images, can be effectively characterized by a single set of kinetic parameters to describe tracer behavior [14].

Over the years, many researchers have explored various methods for obtaining kinetic parameter images. In previous studies, methods for reconstructing kinetic parameter from dynamic PET data can be divided into two main categories: indirect reconstruction methods and direct reconstruction methods [15]. With indirect reconstruction, the focus is on treating image reconstruction and kinetic parameter estimation as separate tasks. Indirect approaches involve initially reconstructing PET emission images for each measurement time point and subsequently estimating the kinetic parameters at each voxel. This approach assumes that the activity distributions are already known as prior information, and the goal is to estimate the kinetic parameters by fitting a mathematical model to the time activity curves (TACs) derived from these activity distributions [16]. Huang *et al.* [17] incorporated a spatial smoothing procedure during the iterations of a nonlinear estimation process at each voxel.

Direct reconstruction integrates image reconstruction and kinetic parameter estimation into a single step, directly estimating kinetic parameter images from raw emission data [18-19]. In 1985, Carson and Lange [20] introduced their research on direct parametric reconstruction using the expectation maximization (EM) algorithm. Inspired by their work, later in 2005, Kamasak *et al.* [14] demonstrated and evaluated a new and efficient method for direct voxel-wise reconstruction of kinetic parameter images using all frames of the PET data. The study states that direct reconstruction methods frequently exhibit superior performance compared to indirect methods, albeit at the expense of increased complexity. In both direct and indirect methods, it is necessary to utilize an appropriate model to fit tracer kinetics and express TACs through mathematical functions. The compartmental model is extensively employed to illustrate the linkage between kinetic parameters and the fundamental physiological processes among these models. Examples of notable works comprise spectral analysis [21], DEPICT [22], and sparse Bayesian learning [23]. Utilizing the compartment model theory, TACs are derived by summing the activity concentration within individual compartments. With a predefined range of basis functions, the fitting process can be regarded as a regression task incorporating biologically plausible sparsity assumptions. This enhances the model's resilience to noise[15].

In recent years, deep learning has shown great potential in medical image fusion[24], image segmentation[25], and image registration[26]. Some researchers have begun combining compartmental models with deep neural networks for kinetic parameter prediction. In generative models, Song et al. [53] proposed a score-based generative framework that formulates both the forward and reverse diffusion processes as stochastic differential equations (SDEs). This framework enables high-quality image synthesis without adversarial training, and introduces Noise Conditional Score Networks (NCSNs) for learning and sampling in high-dimensional spaces. However, the training of generative models only ensures spatial consistency and lacks the preservation of dynamic metabolic information. For convolutional neural networks, attention mechanisms and skip connections are powerful tools for enhancing feature extraction and reducing network information loss. Although skip connections can partially alleviate the loss of fine-grained information, they still rely on heuristic feature fusion. Benetti *et al.* [27] proposed a spatial-temporal self-supervised network (ST-Net) based on the 3D U-Net architecture to unsupervised derive the kinetic parameters of a two-compartment model for $^{18}$F-FDG. Liang *et al.* [28] proposed a kinetic model network (KM-Net) capable of using dynamic PET images from the first 30 minutes to predict kinetic parameters, which are then utilized to reconstruct dynamic PET images for the subsequent 30 minutes. The results of the above two methods indicate that combining the traditional compartment model with deep learning methods can yield favorable outcomes in the prediction of kinetic parameters and dynamic PET images. Specifically, ST-Net mainly focuses on spatio-temporal feature learning using convolutional modules. While effective for capturing local dynamics, it tends to compress high-dimensional information into lower-dimensional feature maps, which may result in the loss of fine-grained temporal variations that are critical for accurate kinetic modeling.KM-Net incorporates kinetic priors more explicitly but relies on conventional CNN architectures, where information may still be discarded across layers. Moreover, both ST-Net and KM-Net primarily optimize for dynamic image reconstruction, with limited emphasis on parametric accuracy.

In many cases, the scanning time required for each frame of dynamic PET images gradually increases, with the entire process taking almost one hour [28]. Long scanning times reduce the comfort of both patients and medical staff. Researchers have developed various techniques to reduce the scanning time for dynamic PET. Some researchers reduce the scanning time of dynamic PET without using deep learning methods. Torizuka *et al*. [29] pointed out that a 30-minute rapid dynamic PET scan can produce interpretation and chamber modeling results similar to a 60-minute dynamic PET scan. In addition, Monden *et al*. [30] demonstrated that shortening the $^{18}$F-FDG brain PET scan time from 60 minutes to 40 minutes had no significant effect on the estimation of kinetic parameters ($k_3$ and $K_i$). Sanaat *et al*. [31] implemented deep learning methods to predict the final 65-minute image from the initial 25-minute PET image, thereby reducing scanning time.

To address this issue, this study proposes a multi-module network (M$^2$-Net) comprising reversible and irreversible modules. This method utilizes the powerful performance of invertible neural networks, which can establish bidirectional mappings between inputs and outputs [32], to construct reversible module. The reversible modules achieve the mapping from earlier frames of dynamic PET images to kinetic parameters. The irreversible modules are constructed based on compartment model theory and parametric image theory, establishing different modality constraints for the output of the reversible modules, thereby improving the accuracy of the network's prediction of kinetic parameter images. Ultimately, the multi-module network can predict high-precision kinetic parameter images from the earlier frames of dynamic PET images and reconstruct dynamic PET images for all frames through kinetic model algorithms, thereby reducing the scan time for dynamic PET images. The digital phantom of the paper can be downloaded from the website: *https://github.com/yqx7150/M2NetPhantom.*

The main contributions of this work are summarized as follows:

***Multi-Module Network****:* This method constructs reversible and irreversible sub-modules to handle bidirectional and unidirectional mappings between inputs and outputs. The reversible module learns the input-output mapping based on an invertible network, while the irreversible module adds constraints to the network using traditional physics-based algorithms. Therefore, multiple modules that integrate learnable-driven and physics-inspired component provide more comprehensive constraints to the network.

***Invertible Network as Reversible Module****:* Invertible networks minimize information loss during input-output transformations to the greatest extent possible. The powerful mapping learning ability of invertible network enables the generation of high-quality intermediate state outputs, thereby ensuring high-quality final outputs from the network.

***Intermediate State Output****:* The kinetic parameter image represents the intermediate state output of the entire network, providing the potential for the subsequent generation of dynamic PET images and parametric images. The presence of the intermediate state output enables the network to be optimized using multimodal information during the training process.

The rest of the manuscript is organized as follows. Section II introduces the theories used in this work, including the invertible network, the 2-tissue compartment model, the Logan plot, and the Patlak plot. Section III gives a detailed introduction to the composition of each module in the proposed network and training objectives of multi-module network. Section IV describes the experimental setup and presents the experimental results. The discussion of the network is stated in Section V. Finally, Section VI summarizes the article.

## 2. Preliminary

### *2.1. Invertible Neural Networks*

In the field of computer vision, an increasing amount of research indicates the importance of convolutional neural networks (CNNs). When appropriately trained, they outperform traditional computer vision-based approaches in tasks such as image generation and image segmentation. Additionally, CNNs can automatically extract features from images through training [28]. However, it has been demonstrated in the literature that conventional CNNs often suffer from the loss of significant depth information in input images [33]. To overcome this limitation, we introduce invertible neural networks (INNs), which mitigate information loss by learning reversible representations under specific conditions [34]. INNs also possess the capability to model inverse scenes [35]. INNs belong to a category of networks that establish bijective mappings between input and output. The network learns a reversible mapping $Y = f(X)$ , with its corresponding inverse mapping as $X = f^{-1}(Y)$. When the input image is $X$, there exists a unique invertible network $f$ such that the output is $Y$ and vice versa. This guarantees full preservation of information during both the forward and reverse transformations.

*2.2. 2-Tissue Compartment Model*

In PET applications, compartment models are typically employed to characterize the absorption, metabolic processes, and elimination of radiotracers within tissue or cells [36]. Given that the majority of researchers have investigated the 2-tissue compartment model (2TCM) and confirmed its efficacy [10], our approach is also based on the 2TCM. The ordinary differential equations (ODEs) of the 2TCM are described as follows:

$$\frac{\mathrm{d}C_1(t)}{\mathrm{d}t} = K_1 C_P(t) - (k_2 + k_3)C_1(t) + k_4 C_2(t) \tag{1}$$

$$\frac{\mathrm{d}C_2(t)}{\mathrm{d}t} = k_3 C_1(t) - k_4 C_2(t) \tag{2}$$

The input function $C_p(t)$ is usually referred to as the radioactivity concentration in blood or plasma. We commonly designate the first compartment as the free compartment, and the second compartment as the binding compartment [37]. Their respective TACs are denoted as $C_1(t)$ and $C_2(t)$. $K_1$, $k_2$, $k_3$, and $k_4$, describe the rate at which the tracer undergoes directional exchanges between various tissue compartments or between the plasma and tissue. Eq. (3) illustrates that the tissue concentration $C_T(t)$ is the sum of the nondisplaceable and binding compartment concentrations [38] and provides the solution for $C_T(t)$:

$$\begin{aligned} C_T(t) &= C_1(t) + C_2(t) \\ &= \frac{K_1}{\beta_2 - \beta_1}\begin{bmatrix} (k_3 + k_4 - \beta_1)e^{-\beta_1 t} \\ +(\beta_1 - k_3 - k_4)e^{-\beta_2 t} \end{bmatrix} \otimes C_P(t) \end{aligned} \tag{3}$$

The specific definitions of $\beta_1$ and $\beta_2$ are shown in [37]. At time frame $k$, the intensity of each voxel in the image can be calculated using the following equation:

$$x_k = \int_{t_s}^{t_e} \left( (1 - V_B)C_T(\tau) + V_B C_B(\tau) \right) e^{-\lambda\tau} \mathrm{d}\tau \tag{4}$$

where $t_s$ and $t_e$ represents the start and end time of the $k$-th frame, respectively. $\lambda$ represents the decay constant of the radiotracer. The constant $V_B$ denotes the volume fraction of a voxel composed of blood. $C_B$ represents the radioactivity concentration of the whole blood.

Combining the above formula, once the four kinetic parameters ($K_1, k_2, k_3, k_4$) of the 2TCM for each voxel are obtained, the entire TACs for each voxel can be derived using these parameters. Once the TACs for each voxel is obtained, dynamic PET images at any desired time can be calculated.

### *2.3. Logan Plot and Patlak Plot*

Once $C_T(t)$ is available, the Logan plot can be drawn based on the following formula [39]:

$$\frac{\int_0^t C_T(\tau)d\tau}{C_T(t)} = K\frac{\int_0^t C_P(\tau)d\tau}{C_T(t)} + b, t > t^* \tag{5}$$

The Logan plot is a graphical analysis method used to evaluate reversible receptor ligands. Based on linear regression, Logan plots are easy to compute. The Logan reference plot describes the relationship between the radioactivity concentration in the target tissue at time $t$, $C_T(t)$, and the radioactivity concentration in the plasma, $C_p(t)$. This means that after the tracer reaches equilibrium at time $t^*$, there is a linear relationship, where the slope of the plot is represented by $K$, and the intercept is represented by $b$.

$$\frac{x(t)}{C_p(t)} = K_i\frac{\int_0^t C_p(\tau)\mathrm{d}\tau}{C_p(t)} + V_0 \tag{6}$$

The Patlak plot can be drawn using the above equation [40]. $K_i$ represents the constant rate of irreversible binding. $V_0$ stands for the distribution volume of the nonspecifically bound tracer in the tissue. $x(t)$ denotes the integrated activity of the tissue up to time t. $C_p(t)$ indicates the plasma concentration of the tracer at time t. For systems with irreversible compartments, this plot will result in a straight line after sufficient equilibration time.

## 3. Proposed Method

### *3.1. Overall of $M^2$-Net*

Based on the description in Section II, obtaining kinetic parameter images makes it possible to acquire both dynamic PET images [37] and parametric images [39-40]. Additionally, we observe that invertible networks can achieve nearly lossless bidirectional mappings between inputs and outputs [41]. Inspired by the above, we propose a multi-module network comprising both reversible and irreversible modules. This network has several characteristics as follows:

Firstly, the network has intermediate state outputs: the kinetic parameter images serve as the intermediate state output, bridging the acquisition of dynamic PET images and parametric images. The intermediate state output, along with the subsequent parametric images and dynamic PET images, adds constraints to the network.

Secondly, the network includes reversible and irreversible modules: the reversible modules achieve bidirectional mappings between dynamic PET images and kinetic parameter images, while the irreversible modules achieve unidirectional mappings from kinetic parameter images to both dynamic PET images and parametric images.

Lastly, the network consists of learnable and non-learnable modules: the reversible module, utilizing invertible network, can learn end-to-end mappings and optimize internal parameters, making it learnable. In contrast, the irreversible modules implement mappings through algorithms based on traditional theories, thereby adding constraint terms, making them non-learnable.

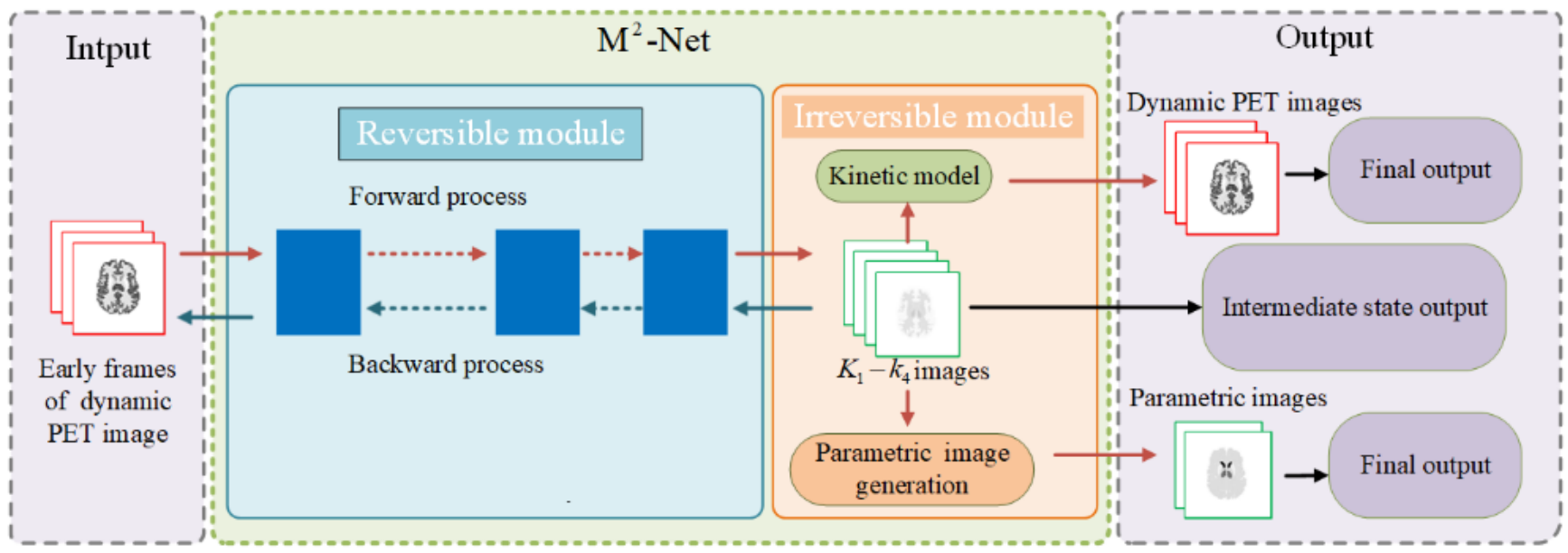

**Fig. 1.** The training process of $M^2$-Net: from left to right are the input, $M^2$-Net and output. $M^2$-Net contains reversible module and irreversible module. The input, consisting of the earlier frames of dynamic PET images, passes through the reversible module to produce the intermediate state output: kinetic parameter images. The kinetic parameter images then pass through the irreversible module to obtain the final output.

The training process of $M^2$-Net is shown in Fig. 1. As shown in the figure: first, the earlier frames of the dynamic PET images are input into the reversible module. Then, the reversible module predicts the kinetic parameter images, which are passed on to the irreversible module to generate the complete dynamic PET images and parametric images. These dynamic PET images and parameter images are the final output of the network. During this process, the network introduces multiple constraint terms to optimize the intermediate state outputs: the forward output of the reversible network, the backward output of the reversible network, and the loss between the generated parametric images and the generated dynamic PET images and their corresponding target images. Notably, the forward and backward processes share the same set of weights. This weight sharing imposes a self-consistency constraint on the model, requiring the mapping and its inverse to be mutually consistent. As shown in Fig. 2, during the testing process, the network predicts the kinetic parameter images, which are then used by the kinetic model algorithm to generate the complete dynamic PET image.

### *3.2. Reversible Module in $M^2$-Net*

Due to the learnability of reversible modules and the generation of intermediate state outputs by the network, the reversible module serves as the core module of the network, as shown in Fig. 1. This section employs an invertible network to implement the reversible module, achieving an accurate mapping from input to intermediate state output. The invertible network's detailed architecture is illustrated in Fig. 3. It is composed of multiple invertible blocks, each comprising invertible convolution and affine coupling layers.

Initially, the input image is split into two halves along the channel dimension. The transformations $s$ , $t$ , and $r$ are structured similarly to dense blocks, consisting of eight 2D convolution layers with a filter size of 3 × 3. Each layer learns a new set of feature maps from the preceding layer. The first four convolutional layers have a receptive field size of 1 × 1 and a stride of 2, followed by a rectified linear unit (ReLU). The final layer is a 3 × 3 convolutions without ReLU. Incorporating Leaky ReLU layers aims to prevent overfitting to the training set [42] and enhance nonlinearity, thereby improving the invertible network's capability to separate dual-tracer PET images. During the forward process, the input image undergoes transformations via a sequence of bijective functions, resulting in output images. Our research aims to find a one-to-one function that effectively maps data points from dynamic PET images to different kinetic parameters images. Inspired by [41], we introduced an affine coupling layer to ensure bidirectional mapping based on a single network.

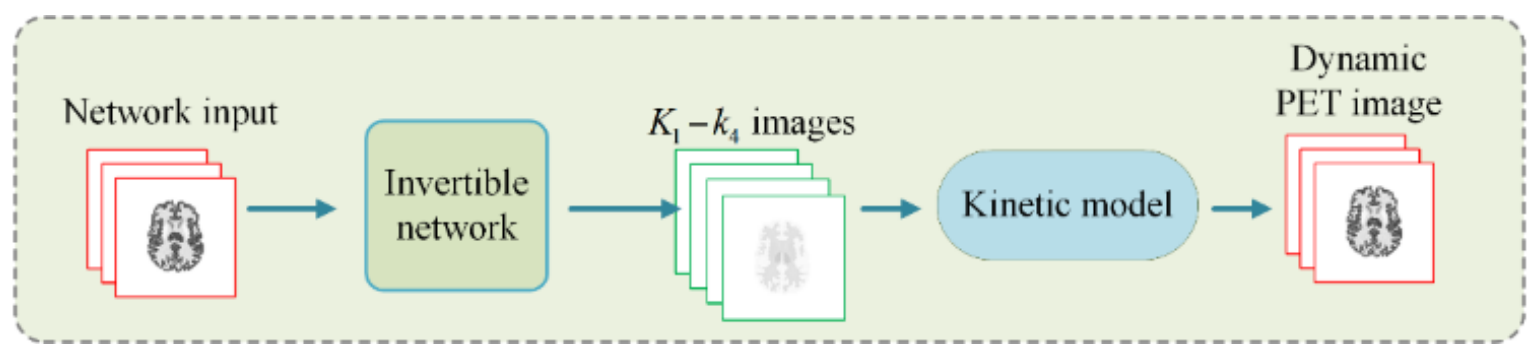

**Fig. 2.** The testing process of $M^2$-Net. The input to the reversible network consists of dynamic PET images from earlier frames, and the output is the kinetic parameter images. The kinetic parameter images are then processed by the kinetic model algorithm to obtain the complete dynamic PET images.

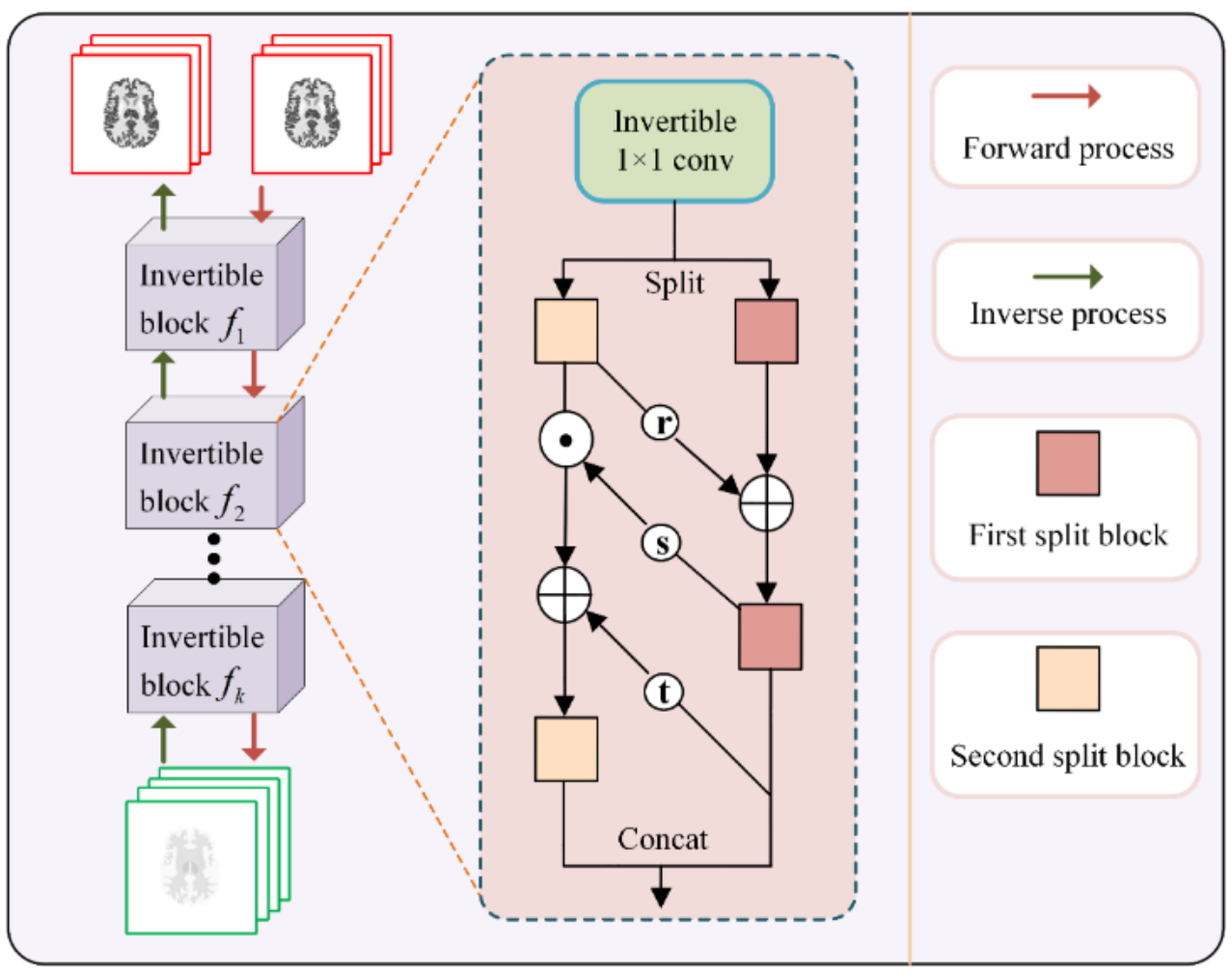

**Fig. 3.** Pipeline of proposed invertible network and the details of the invertible block. Each invertible block consists of invertible $1 \times 1$ convolution and affine coupling layers.

Specifically, our invertible network is constructed by amalgamating a sequence of reversible and tractable viable bijective functions $\{f_i\}_{i=0}^{k}$, $i.e.$, $f = f_0 \circ f_1 \circ f_2 \circ \cdots \circ f_k$. For a given data sample, we can achieve the mutual transformation between input data and output data using the following formula:

$$y = f_0 \circ f_1 \circ f_2 \circ \cdots \circ f_k(x) \tag{7}$$

$$x = f_k^{-1} \circ f_{k-1}^{-1} \circ \cdots \circ f_0^{-1}(y) \tag{8}$$

This study implements the bijective model $f_i$ using affine coupling layers. Within each affine coupling layer, with a $D$-dimensional input m and $d < D$, the output $n$ is computed as:

$$n_{1:d} = m_{1:d} \tag{9}$$

$$n_{d+1:D} = m_{d+1:D} \odot \exp\big(s(m_{1:d})\big) + t(m_{1:d}) \tag{10}$$

The functions $s$ and $t$ denote scaling and translation operations from $R^D \rightarrow R^{D-d}$, respectively, and $\odot$ represents the Hadamard product. Take note that the scale and translation functions are not necessarily invertible, hence, we realize them through neural networks.

As mentioned in [41], the coupling layer maintains some input channels unchanged, limiting the representation learning capability of the architecture. To address this issue, we improve the coupling layer by:

$$n_{1:d} = m_{1:d} + r(m_{d+1:D}) \tag{11}$$

Here, $r$ can be any function from $R^D \rightarrow R^{D-d}$. The inverse step can be easily obtained through the following equation:

$$m_{d+1:D} = \left(n_{d+1:D} - t(n_{1:d})\right) \odot \exp\left(-s(n_{1:d})\right) \tag{12}$$

$$m_{1:d} = n_{1:d} - r(m_{d+1:D}) \tag{13}$$

Next, we employ the invertible $1 \times 1$ convolution introduced in [43] as the adaptable permutation function to invert the channel order for the subsequent affine coupling layer.

### *3.3. Irreversible Module in M$^2$-Net*

As described in the 2-tissue compartment model section of Section II, once the kinetic parameters of each voxel are obtained, the intensity value of each voxel at different time frames can be calculated, resulting in a dynamic PET image. As shown in the kinetic model algorithm in Fig. 1, this study designs the algorithm to convert kinetic parameter images into dynamic PET images. The algorithm first obtains the kinetic parameter images generated by the reversible module and then applies Eq. (3) and (4) to calculate the value of each voxel in each frame, thus obtaining the dynamic PET images.

| **Algorithm 1: Parametric image generation in irreversible module** |
| --- |
| **Logan plot** |
| **1:** Obtain $\mathrm{K}_1, \mathrm{k}_2, \mathrm{k}_3, \mathrm{k}_4$ generated by invertible network, load $C_P(t)$ |
| **2:** Compute $Pred_C_T(t)$ by Eq. (3) |

**3:** Load target kinetic parameter images $\mathrm{K_1, k_2, k_3, k_4}$

**4:** Compute $C_T(t)$ by Eq. (3)

**5**: $x = \frac{\int_0^t C_P(\tau)d\tau}{C_T(t)}, y = \frac{\int_0^t Pred_C_T(\tau)d\tau}{Pred_C_T(t)}$

**6:** Perform a linear fit according to the equation $y = kx + b$ to compute pred_k and pred_b

**Patlak plot**

**1:** Obtain $\mathrm{K_1, k_2, k_3, k_4}$ generated by invertible network

**2:** Obtain the dynamic PET image Pred_x by Eq. (3) and (4), load $\mathrm{C_P(t)}$

**3**: $x = \frac{\int_0^t C_p(\tau)d\tau}{C_p(t)}, y = \frac{Pred_x}{C_p(t)}$

**4:** Perform a linear fit according to the equation $y = K_i x + V_0$ to compute pred_$\mathrm{K_i}$ and pred_$V_0$

Additionally, according to the Logan plot theory and Patlak plot theory in Section II, Once the kinetic compartmental parameter images are obtained, parametric images of graphical analysis are obtained. In Fig. 1, the dynamic PET image and the parametric image are the final outputs of the network. They are also used to compute the loss with the corresponding target images, adding constraints to the intermediate state outputs.

**Algorithm 1** specifically details how the parametric images are generated. We have designed generation algorithms for both the Logan plot and the Patlak plot to be applicable for training with reversible and irreversible tracers in different data scenarios. The establishment of this algorithm uses Eq. (3), (4), (5), and (6). During the process of performing linear regression to fit the parametric image, we use data corresponding to the last 10 frames.

### *3.4. Training Objectives in M$^2$-Net*

To guarantee the quality of generated images, researchers frequently utilize multi-component cost functions to refine the network. For instance, MM-Synthesis [44] utilized a cost function comprising three distinct components to train a convolutional network model for MRI synthesis. FGEAN [45] integrated both pixel-wise intensity loss and gradient information loss. Hernández *et al.* [46]

trained a residual 3D convolutional network for enhancing the sinogram of small animal PET images by using a loss function that combines a fidelity term and a regularization term. These studies use multiple loss functions to ensure the effectiveness of network performance. This study employs a multi-component loss function to optimize the network. The training objective of the network is as follows:

$$\mathcal{L}_{total} = \mathcal{L}_1 + \lambda_1 \mathcal{L}_2 + \lambda_2 \mathcal{L}_3 + \lambda_3 \mathcal{L}_4 \quad (14)$$

In Fig. 1, the losses between the outputs of the backward and forward processes of the reversible module and their respective target images are calculated, resulting in $\mathcal{L}_1$ and $\mathcal{L}_2$. $\mathcal{L}_3$ and $\mathcal{L}_4$ are the losses computed from the final output of the network. The kinetic parameter images obtained from the output of the reversible module are processed by the kinetic model algorithm and parametric image generation algorithm of the irreversible module to obtain the dynamic PET images and parametric images, respectively. $\mathcal{L}_3$ represents the loss between the obtained dynamic PET images and the target dynamic PET images. $\mathcal{L}_4$ represents the loss between the obtained parametric images and the target parametric images. The aforementioned losses, including the forward and backward losses of the reversible modules, the dynamic PET image and parametric image losses of the irreversible modules, collectively contribute to the optimization of the kinetic parameter images during the network learning process through the combined influence of multiple modalities of information. Hyper-parameters $\lambda_1$, $\lambda_2$ and $\lambda_3$ are used to balance the loss functions. This method uses mean square error (MSE) loss as the loss function. Under the combined influence of multiple losses, the network ensures the accuracy of predicted kinetic parameter images.

## 4. Experiments

This study uses ST-Net [29] and KM-Net [28] as benchmark methods to evaluate the performance of our proposed method. In the experiment, for the irreversible tracer $^{18}$F-FDG and the reversible tracer $^{11}$C-FMZ, the $\mathcal{L}_4$ of M$^2$-Net is adjusted to the Patlak plot loss and Logan plot loss, respectively, as described in **Algorithm 1**. When using reversible tracers, the $\mathcal{L}_2$ of the KM-Net has been modified from the original Patlak plot form to the Logan plot form to adapt. To ensure a uniform comparison, we applied the kinetic model algorithm to convert the kinetic parameter images predicted by ST-Net into dynamic PET images during its testing.

*4.1. Experimental Setup*

***Simulation Datasets:*** As illustrated in Fig. 4, we utilized two distinct phantoms for simulating dynamic PET images: the Zubal brain phantom and the Zubal thorax phantom. The size of each layer in both phantoms is $128 \times 128$. The voxel size is $2.2 \times 2.2 \times 1.4$ mm$^3$. Random motion fields were applied to augment the data for each phantom. For each phantom, we delineated five and three regions of interest (ROI) respectively.

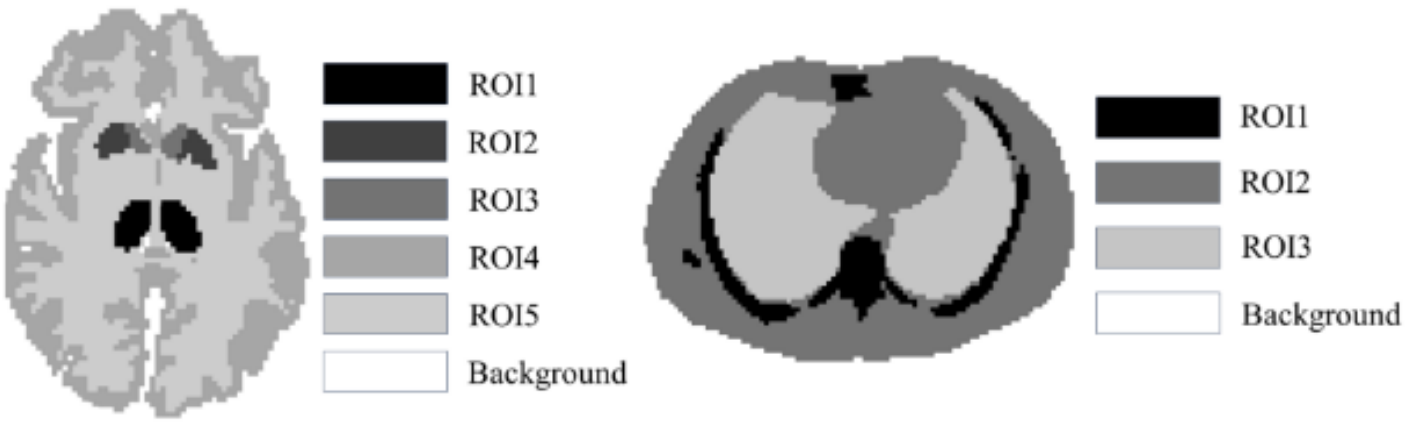

**Fig. 4.** Two phantoms used in the simulation experiment: (a) Zubal brain phantom (b) Zubal thorax phantom. The two phantoms have 5 and 3 ROIs, respectively.

The generation method for the network input and the target images corresponding to the network output in the simulation experiment is as follows: The input function $\mathrm{C_p(t)}$ for $^{11}$C-FMZ and $^{18}$F-FDG is modelled according to the reference [47-48]. Physiological variations were simulated by Gaussian randomization of the kinetic parameters ($K_1, k_2, k_3, k_4$). The mean values of these parameters were set according to references [48-51], with a standard deviation of 20% of the mean. After setting up the phantom, parameters, and scanning protocol as described above, tissue concentrations $\mathrm{C_T(t)}$ were obtained by solving the ODEs using MATLAB tools, as described in Section II. Using Eq. (4), noise-free dynamic activity images for all voxels were generated, with image dimensions of $128 \times 128 \times 18$. Noise-free sinogram data for each frame was generated by forward-projecting the corresponding ground-truth images, and then Poisson noise was added to simulate different count levels. The dynamic sinograms were reconstructed into noisy dynamic activity images using the OSEM[52] algorithm with 6 iterations and 5 subsets. Consistent with the phantom and simulated noise-free activity images, the voxel dimensions of the reconstructed images were 2.2 $\times$ 2.2 $\times$ 1.4 $\mathrm{mm}^3$. The kinetic parameter values used in the aforementioned process were filled into each ROI to obtain corresponding kinetic parameter images. Specifically, for different tracers and phantoms, we calculated that the dose of the Zubal brain phantom based on FDG based on different ROIs is [3.0534: 1.9219: 1.1160: 3.4030: 2.8997], the dose of the Zubal thorax phantom based on FDG based on different ROIs is [1.6765: 5.3815: 6.4493], and the dose of the Zubal brain phantom based on FMZ based on different ROIs is [1.7414: 2.0374: 1.0804: 3.4119: 0.4763]. Finally, the last 10 frames of the dynamic PET images were fitted using the linear kinetic analysis model to produce the slope and intercept images.

**Table 1**

Settings of the simulation experiments

| Task | Tracer | Phantom | Scanning time | Scanning interval | Noise level |
|---|---|---|---|---|---|
| 1 | $^{18}$F-FDG | Zubal brain | 60 min | 4 × 30 s + 4 × 120 s + 10 × 300 s | 0.1 |
| 2 | $^{11}$C-FMZ | Zubal brain | 60 min | 4 × 30 s + 4 × 120 s + 10 × 300 s | 0.1 |
| 3 | $^{18}$F-FDG | Zubal thorax | 60 min | 4 × 30 s + 4 × 120 s + 10 × 300 s | 0.1 |
| 4 | $^{18}$F-FDG | Zubal brain | 60 min | 4 × 30 s + 4 × 120 s + 10 × 300 s | 0.2 |
| 5 | $^{18}$F-FDG | Zubal brain | 60 min | 4 × 30 s + 4 × 120 s + 10 × 300 s | 0.3 |

In this section, a total of five simulated datasets were generated, labelled as Task 1 to Task 5. Task 1 investigates the performance of the three methods with the tracer $^{18}$F-FDG and the Zubal brain phantom. Task 2 examines the performance of the three methods with the tracer $^{11}$C-FMZ and the Zubal brain phantom. Task 3 explores the performance of the three methods with the tracer $^{18}$F-FDG and the Zubal thorax phantom. Task 1, Task 4 and Task 5 focus on exploring the performance of three methods on data with different levels of noise. The noise level refers to the statistical variation of the acquired PET counts. These tasks are used to study the robustness of different methods under varying tracers, phantoms and noise levels Specifically, Poisson noise was introduced at three relative levels (0.1, 0.2, and 0.3), corresponding approximately to average total analog count levels of $8.3 \times 10^5$, $4.1 \times 10^5$ and $2.7 \times 10^5$ events per frame, respectively. The detailed settings, including total scanning time, intervals, tracers, phantoms, and noise levels, are listed in Table 1.

***Clinical Datasets:*** The research was approved by the Ethics Committee at Renji Hospital, Shanghai Jiao Tong University School of Medicine(lRB). Informed consents from all human subjects involved in the research have been obtained. The clinical dataset in this section was obtained from Renji Hospital using the uEXPLORER machine provided by United Imaging Healthcare (UIH). The dataset consists of brain scans from 10 patients, with $^{18}$F-FDG used as the tracer for the scans. For each patient, 4 layers of brain data were selected to form the dataset. The dynamic PET data spans a total of 60 minutes, comprising 18 frames of $128 \times 128$ images: $4 \times 30$ s, $4 \times 120$ s and $10 \times 300$ s. The scanning interval of the clinical data is the same as that of the simulated data set in Task 1, and the same tracer was used.

***Parameter Configuration:*** For the simulation study, 180 pairs of data were used for training and 20 pairs of data were used for testing. During training, the input for all three methods is set as 12 frames of noisy dynamic PET images corresponding to the first 30 minutes of scanning. During testing, the input for all three methods is also 12 frames of noisy dynamic PET images from the first 30 minutes of scanning, with the output being the complete dynamic PET images. These consistent settings for the three methods ensure fairness in the experiments. All data used as input in the experiments were normalized. The Adam optimizer is employed to train all networks in our experiments. We train the proposed model for 300 epochs with a batch size of 1. Initially, the learnable component of the model consists of k=8 invertible blocks, each constructed with affine coupling layers, with both input and output channel numbers being 24, the learning rate is set to 0.0001 for the first 50 epochs, and then it is halved every 50 epochs. This gradual reduction in the learning rate persists as the number of epochs increases. The network training parameters for the other two comparative methods are designed according to their respective papers. All experiments are conducted using a customized version of PyTorch on hardware consisting of an NVIDIA Tesla V100.

*4.2. Simulation Results*

The experiments compare the last 6 frames of dynamic PET images reconstructed from kinetic parameter images ($K_1, k_2, k_3, k_4$) predicted by three different methods. This study employs mean square error, structural similarity (SSIM) and peak signal to noise ratio (PSNR) as the primary metrics to assess experimental results.

$$\text{Bias} = \frac{1}{N}\sum_{\text{i}=1}^{N}\frac{|x_\text{i} - \hat{x}_\text{i}|}{x_\text{i}} \tag{15}$$

$$\text{Variance} = \frac{1}{N}\sum_{\text{i}=1}^{N}\left(\frac{x_\text{i} - \overline{x_N}}{x_\text{i}}\right)^2 \tag{16}$$

Additionally, we compute the bias and variance using Eq. (15) and (16). Here, $N$ represents the total number of pixels in the ROI, $x_\text{i}$ and $\hat{x}_\text{i}$ denote the ground truth and prediction results of the i-th pixel respectively, and $\overline{x_N}$ represents the average predicted value of the pixels in the ROI.

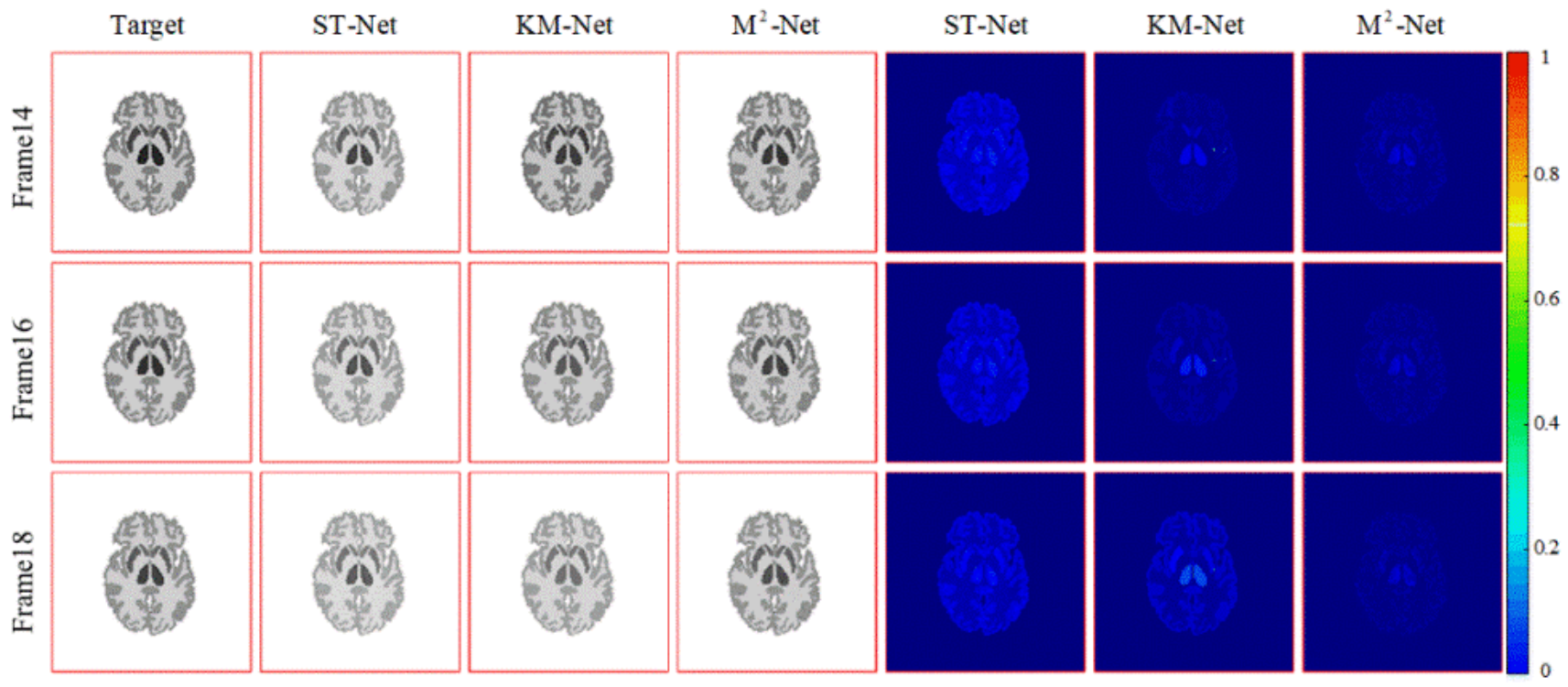

**Fig. 5.** The test results of ST-Net, KM-Net, and M²-Net with $^{18}$F-FDG as the tracer are shown. From left to right: the target images, the prediction results of ST-Net, KM-Net and M²-Net, and the error maps of the predictions by ST-Net, KM-Net, and M²-Net. From top to bottom: the display of test results for frames 14, 16, and 18.

***Tests of $^{18}$F-FDG as the Tracer:*** This section evaluates the performance of the proposed method and two other methods when the tracer for the simulation data is $^{18}$F-FDG. The experiments in this section correspond to Task 1 in Table 1. The quantitative results of the experiments are shown in Table 2. From Table 2, it can be seen that the proposed method outperforms the comparison methods in dynamic PET image prediction. Fig. 5 presents the prediction results and error maps for the last 30 minutes of dynamic PET images by three methods, with images selected from frames 14, 16, and 18.

From the images, it can be seen that the dynamic PET images predicted by the proposed method are closer to the target images. The error maps show that, overall, the $M^2$-Net method has the smallest error in the prediction results, followed by the KM-Net, while the ST-Net has the largest error. In summary, the proposed method demonstrates good predictive performance under the $^{18}$F-FDG tracer.

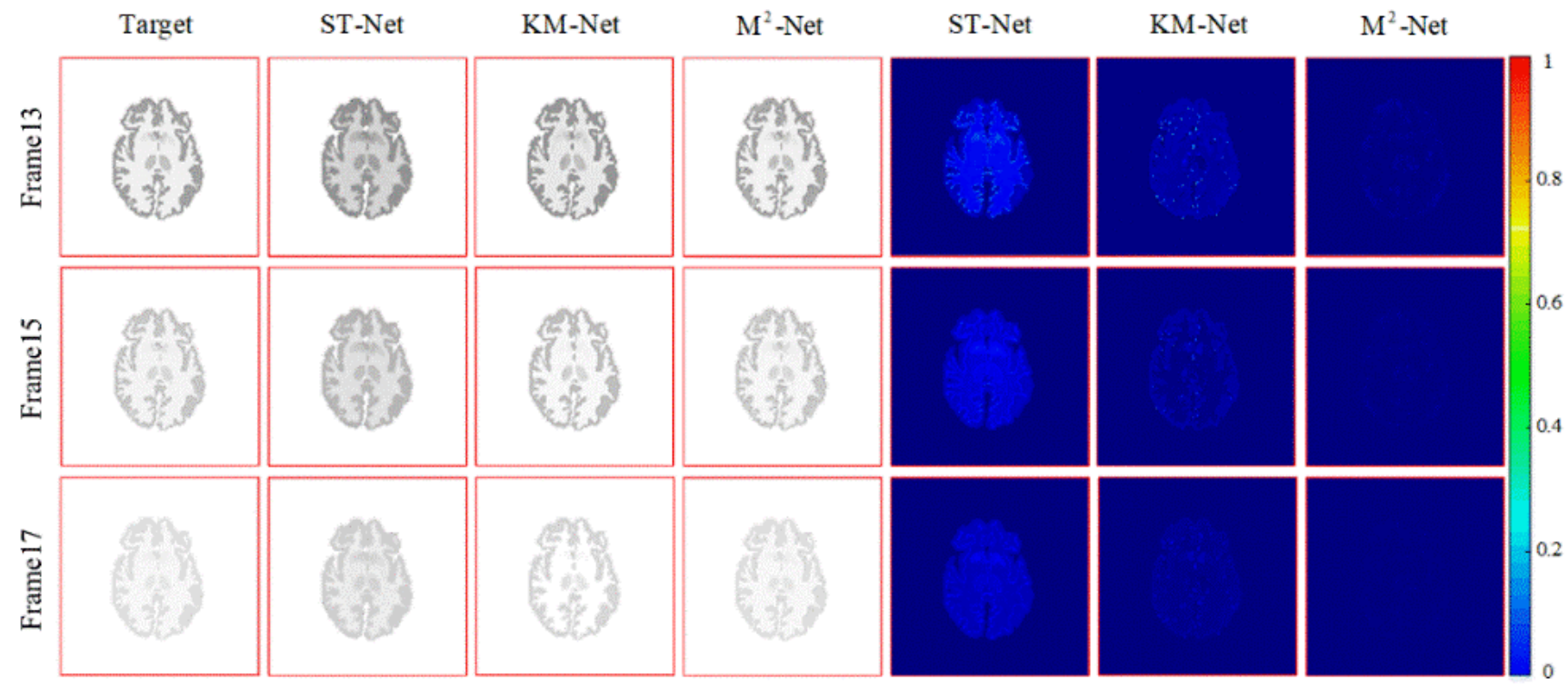

**Fig. 6.** The test results of ST-Net, KM-Net, and $M^2$-Net with $^{11}$C-FMZ as the tracer are shown. From left to right: the target images, the prediction results of ST-Net, KM-Net and $M^2$-Net, and the error maps of the predictions by ST-Net, KM-Net, and $M^2$-Net. From top to bottom: the display of test results for frames 13, 15, and 17.

***Tests of $^{11}$C-FMZ as the Tracer:*** This section explores the performance of the three methods when $^{11}$C-FMZ is used as the tracer. The experiments in this section correspond to Task 2 in Table 1. The test result images for Task 2 are shown in Fig. 6, with specific metrics listed in Table 2. As seen from Table 2, when the tracer for the simulation data is $^{11}$C-FMZ, the PSNR, SSIM, and MSE metrics for $M^2$-Net's predictions are the best among the three methods. Overall, except for SSIM, the metric results of KM-Net are higher than those of ST-Net, achieving higher PSNR values and lower MSE. Fig. 6 presents the prediction results and error maps for the last 30 minutes of dynamic PET images by $M^2$-Net,

ST-Net, and KM-Net. From the prediction results and corresponding residual maps, it is evident that M$^2$-Net provides more accurate dynamic PET images with the smallest error. In contrast, ST-Net and KM-Net introduce larger noise in their predictions.

Fig. 7 shows the kinetic parameter images predicted by M$^2$-Net and the corresponding target images. From the images, it can be seen that the predicted kinetic parameter images are very close to the target images. The predicted kinetic parameter images not only accurately restore the shape of each ROI but also accurately reflect the intensity within each ROI.

Fig. 8 shows the $K$ images of Logan plots generated by linear fitting of the last 10 frames from the predicted images of the three methods, along with their corresponding target images and residual images. From the residual images, it is evident that our method provides a more accurate fit for the $K$ images compared to the competing methods. Table 3 presents the corresponding PSNR, SSIM, and MSE metrics, which further demonstrate that our method achieves more accurate K image fitting than the competing methods.

**Table 2**

Quantitative results of Task 1to Task 5, Including MSE, PSNR, and SSIM. Lower MSE, higher SSIM, and greater PSNR are highlighted in bold.

| TASK | Method | Index | | |
|---|---|---|---|---|
| | | PSNR | SSIM | MSE |
| 1 | M$^2$-Net | **40.72** | **0.9968** | **0.0001** |
| | KM-Net | 32.74 | 0.9543 | 0.0006 |
| | ST-Net | 29.44 | 0.9819 | 0.0012 |
| 2 | M$^2$-Net | **43.08** | **0.9935** | **0.0001** |
| | KM-Net | 34.28 | 0.8297 | 0.0004 |
| | ST-Net | 30.23 | 0.9376 | 0.0012 |
| 3 | M$^2$-Net | **36.08** | **0.9767** | **0.0003** |
| | KM-Net | 27.19 | 0.9614 | 0.0021 |
| | ST-Net | 21.07 | 0.9214 | 0.0079 |
| 4 | M$^2$-Net | **37.18** | **0.9918** | **0.0002** |
| | KM-Net | 32.37 | 0.9799 | 0.0006 |
| | ST-Net | 23.48 | 0.9222 | 0.0048 |
| 5 | M$^2$-Net | **35.46** | 0.9873 | **0.0003** |
| | KM-Net | 32.1 | **0.9888** | 0.0006 |
| | ST-Net | 22.32 | 0.9062 | 0.0064 |

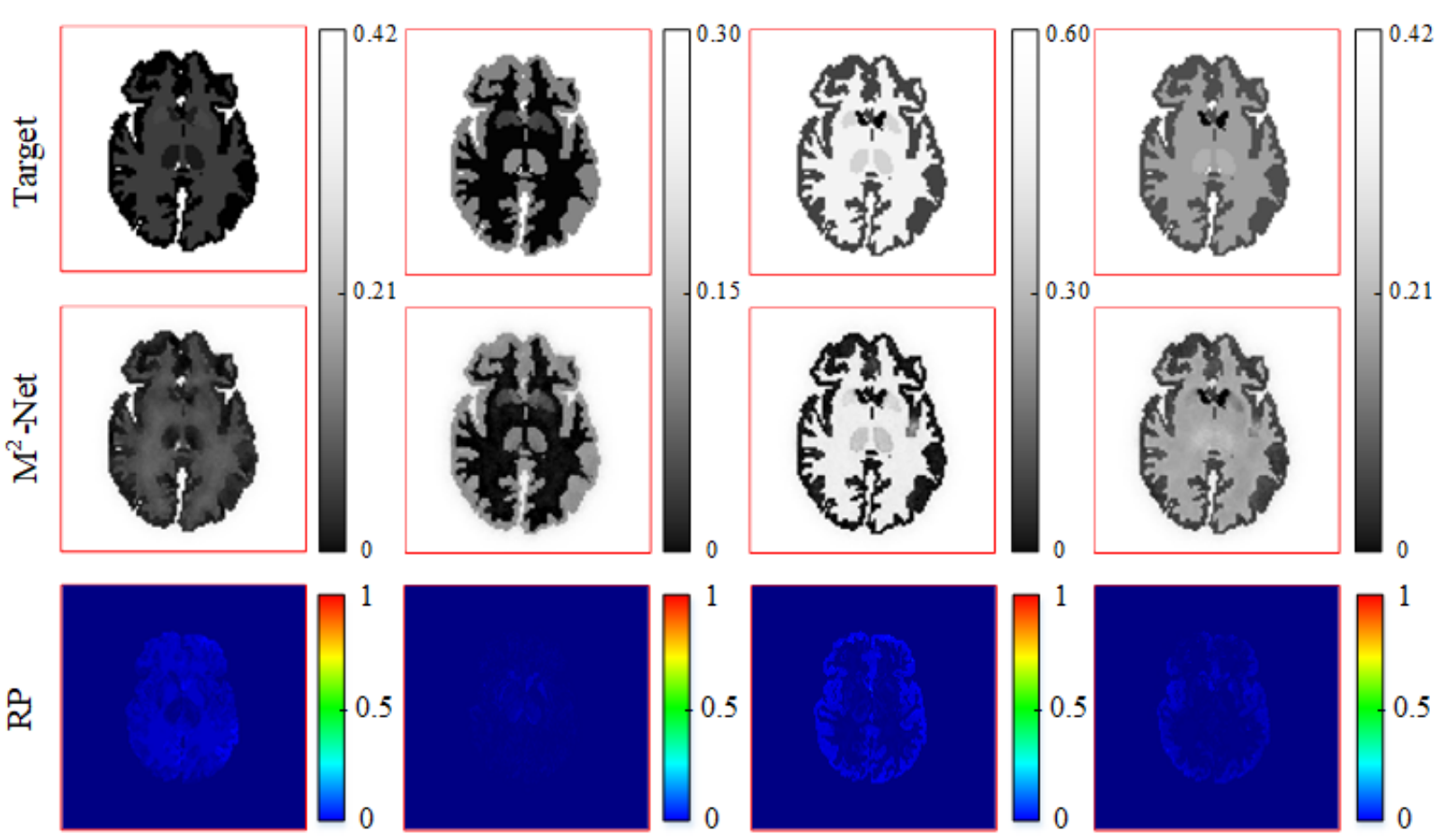

**Fig. 7.** The prediction results of kinetic parameter images by M²-Net are shown as follows: from left to right are the images of $K_1, k_2, k_3$ and $k_4$. From top to bottom are the target images, the images predicted and the residual plot by M²-Net.

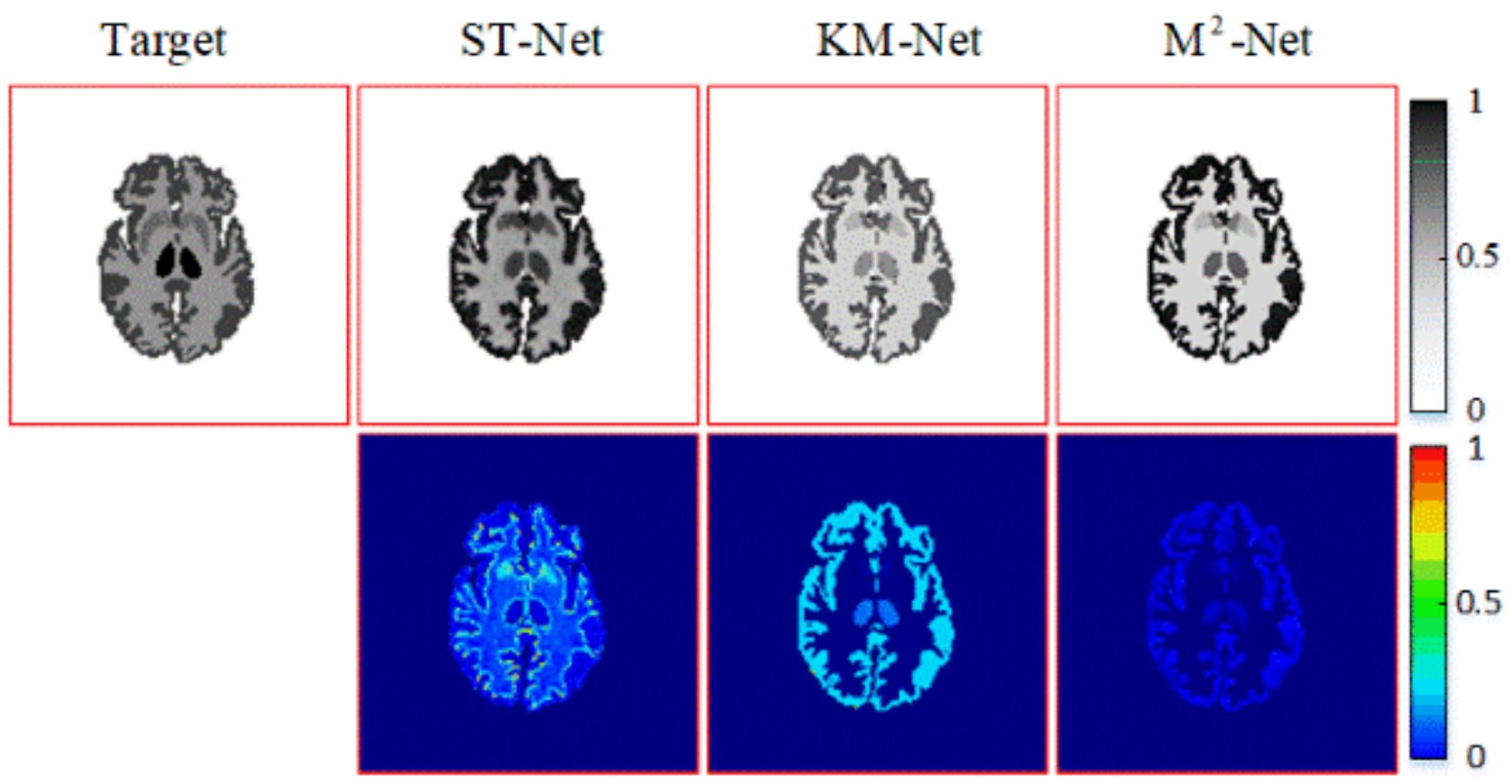

**Fig. 8.** The $K$ image results of ST-Net, KM-Net, and M²-Net using $^{11}$C-FMZ as the tracer are shown. From left to right: the target images, the prediction results of ST-Net, KM-Net and M²-Net, followed by the error maps of the predictions by ST-Net, KM-Net, and M²-Net.

**Table 3**

$K$ Image results of Task 2, including MSE, PSNR, and SSIM. Lower MSE, higher SSIM, and greater PSNR are highlighted in bold.

| Index | ST-Net | KM-Net | M$^2$-Net |
|---|---|---|---|
| PSNR | 20.85 | 23.41 | **30.49** |
| SSIM | 0.9247 | 0.9710 | **0.9960** |
| MSE | 0.0089 | 0.0053 | **0.0009** |

***Tests of Zubal Thorax as Phantom:*** This section explores the performance of the three methods when the Zubal thorax phantom is used. This corresponds to Task 3 in Table 1. The quantitative results for Task 3 are shown in Table 2. When the phantom for the simulation data is changed to the Zubal thorax, $M^2$-Net's prediction results still exhibit better PSNR, SSIM, and MSE metrics. KM-Net's metrics are worse than those of $M^2$-Net, and ST-Net has the worst quantitative metrics.

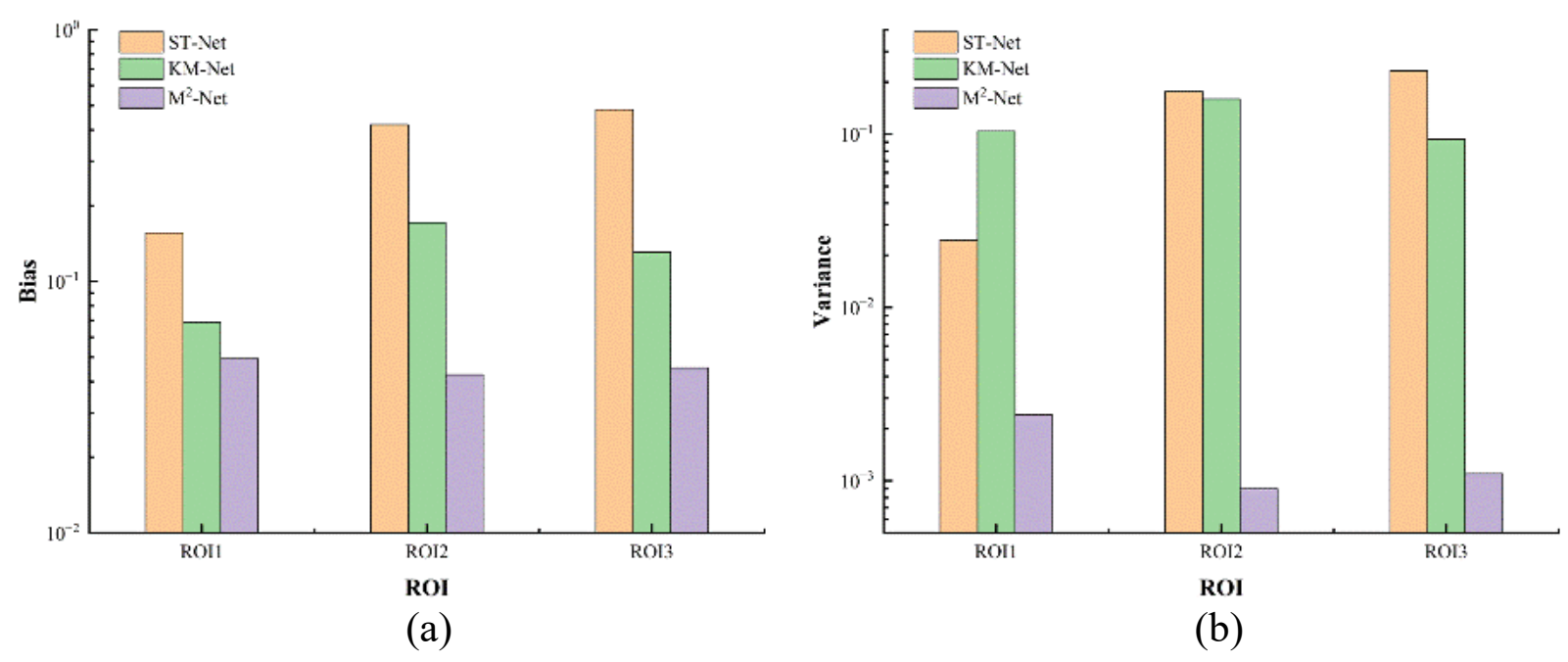

**Fig. 9.** Bias and variance result images for the 13th frame of thorax test data predictions: (a) bias result image and (b) variance result image.

We applied formulas (15) and (16) to compute the bias and variance for the prediction results of the three methods, and the resulting images are shown in Fig. 9. The y-axis in the figures is plotted on a logarithmic scale with base 10. It can be seen from the figures that predictions of $M^2$-Net have the lowest bias and variance, followed by KM-Net, while ST-Net has the highest bias and variance. The tests conducted on the above experiments evaluated the performance of different methods under various data conditions, including changes in tracers for simulated data and variations in phantoms. The results indicate that, compared to the other two methods, $M^2$-Net maintains better performance and demonstrates greater robustness under changes in tracers and phantom.

***Tests under data with different levels of noise:*** Task 1, 4, and 5 in Table 1 study the performance of different methods by setting the noise level of the input data to 0.1, 0.2, and 0.3. The corresponding quantitative results of the experiments are shown in Table 2. From the table, it can be seen that as the noise level of the input data increases, the overall predictive performance of the three methods decreases, but the proposed method still has good performance.

Fig. 10 shows the bias and variance boxplots of three methods for all samples in the test set under different noise levels. It can be clearly seen from the figure that under different levels of noise, the prediction results of M$^2$-Net method have lower bias and variance compared to the other two methods, and the distribution is more concentrated. The bias and variance of KM-Net prediction results are second to M$^2$-Net. The prediction results of the ST-Net method have the largest deviation and variance among the three methods.

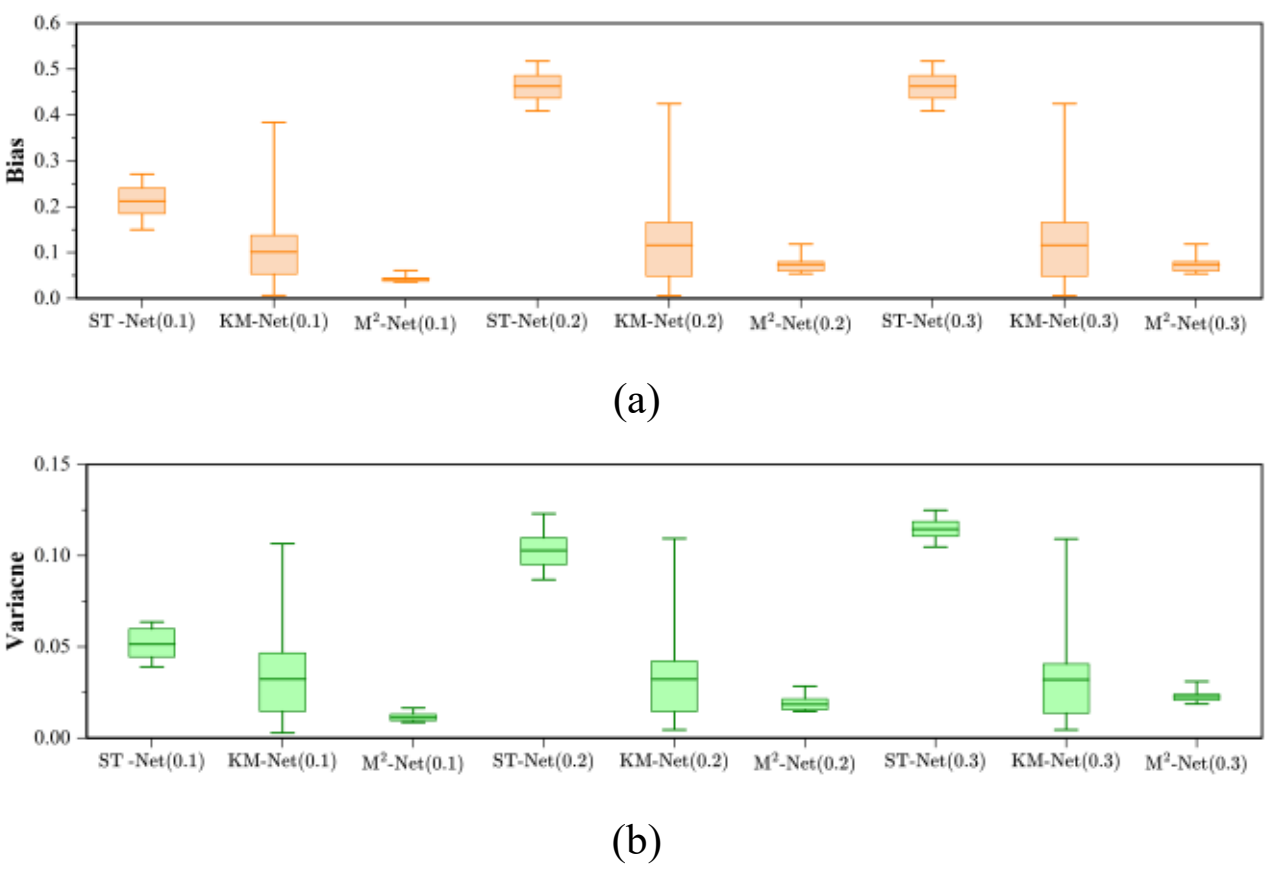

**Fig. 10.** Bias and variance box plots of $^{18}$F-FDG tracer in the subsequent frames predicted by ST-Net, KM-Net, and M²-Net under three different noise levels (a) bias box plot and (b) variance box plot.

### *4.3. Generalization Study*

**Table 4**

Quantitative results of generalization study experiments, Including MSE, PSNR, and SSIM. Lower MSE, higher SSIM, and greater PSNR are highlighted in bold.

| Index | ST-Net | KM-Net | M$^2$-Net |
| --- | --- | --- | --- |
| PSNR | 27.97 | 25.66 | **28.00** |
| SSIM | 0.9397 | 0.9426 | **0.9434** |
| MSE | 0.0030 | 0.0030 | **0.0022** |

This section of the experiment conducted a generalization study using clinical human data. The experiment tested the models trained in Task 1 using three different methods on clinical human data. Table 4 presents the quantitative results of the generalization experiment. The metrics implies that M$^2$-Net achieves the best PSNR, SSIM, and MSE among the three methods. Additionally, the PSNR and SSIM of ST-Net are quite close to those of the $M^2$-Net method.

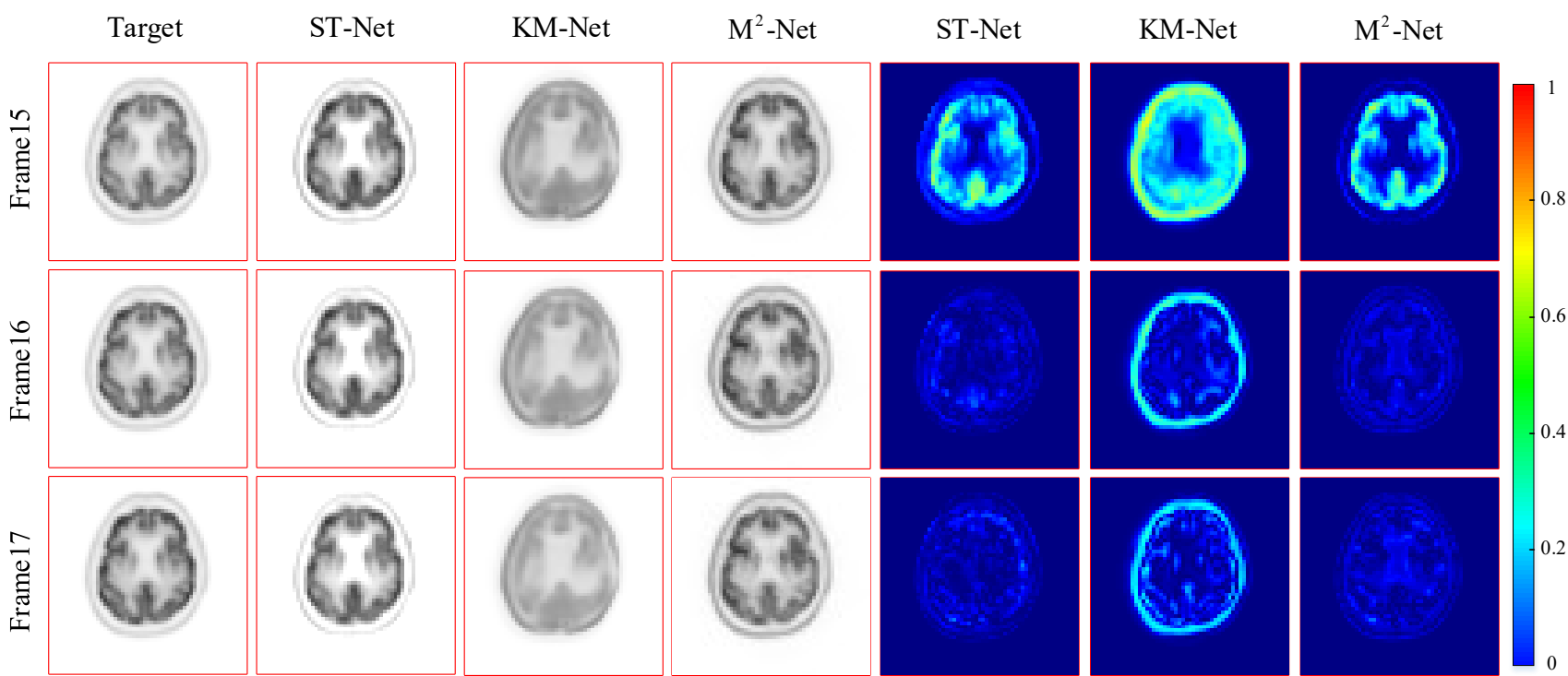

**Fig. 11.** Experimental results with Clinical data: including the 15th, 16th, and 17th frames of the target, as well as the prediction results from three methods. None of the three methods were retrained on clinical data; instead, the models trained in Task 1 were used directly for testing.

The PSNR of KM-Net is comparatively worse than that of $M^2$-Net and ST-Net. KM-Net achieves an SSIM value very close to that of the $M^2$-Net method, and its MSE is the same as that of ST-Net. Fig. 11 displays frames 15 to 17 of the dynamic PET prediction results and the corresponding target images for the three methods. From the images, it can be observed that $M^2$-Net has the best restoration of the target images, including the intensity and specific shapes of different regions. ST-Net also performs well in restoring the shapes of different regions, but there are some deviations in the intensity restoration. KM-Net manages to restore the shapes of most regions but performs poorly in restoring details and shows significant deviations in intensity restoration.

Fig. 12 shows boxplots of the bias and variance of the three methods for all samples in the real data test set. The figure clearly shows that, using the real data, the $M^2$-Net method achieves lower bias than the other two methods, while its variance is comparable to that of ST-Net and significantly better than that of KM-Net. Fig. 13 plots the profile comparison results of the three methods under the 17th frame prediction results of clinical data. It can be observed that the prediction results of $M^2$-Net are closer to the target image, maintaining smaller intensity deviations in most pixels. ST-Net maintains a small intensity deviation over certain pixel intervals. In contrast, KM-Net exhibits larger intensity deviations in most pixels. The clinical data test results demonstrate that the proposed method has good generalization. Moreover, the results are visually rated by an experienced nuclear medicine expert, resulting in a conclusion that aligns with the numeric metrics. For the test dataset out of the data distribution, the proposed model is more stable than the other two methods, the model can generalize well without additional fine-tuning. In future work, once larger datasets are available, we plan to use transfer learning or domain adaptation strategies to integrate real data, which will further improve robustness and clinical applicability.

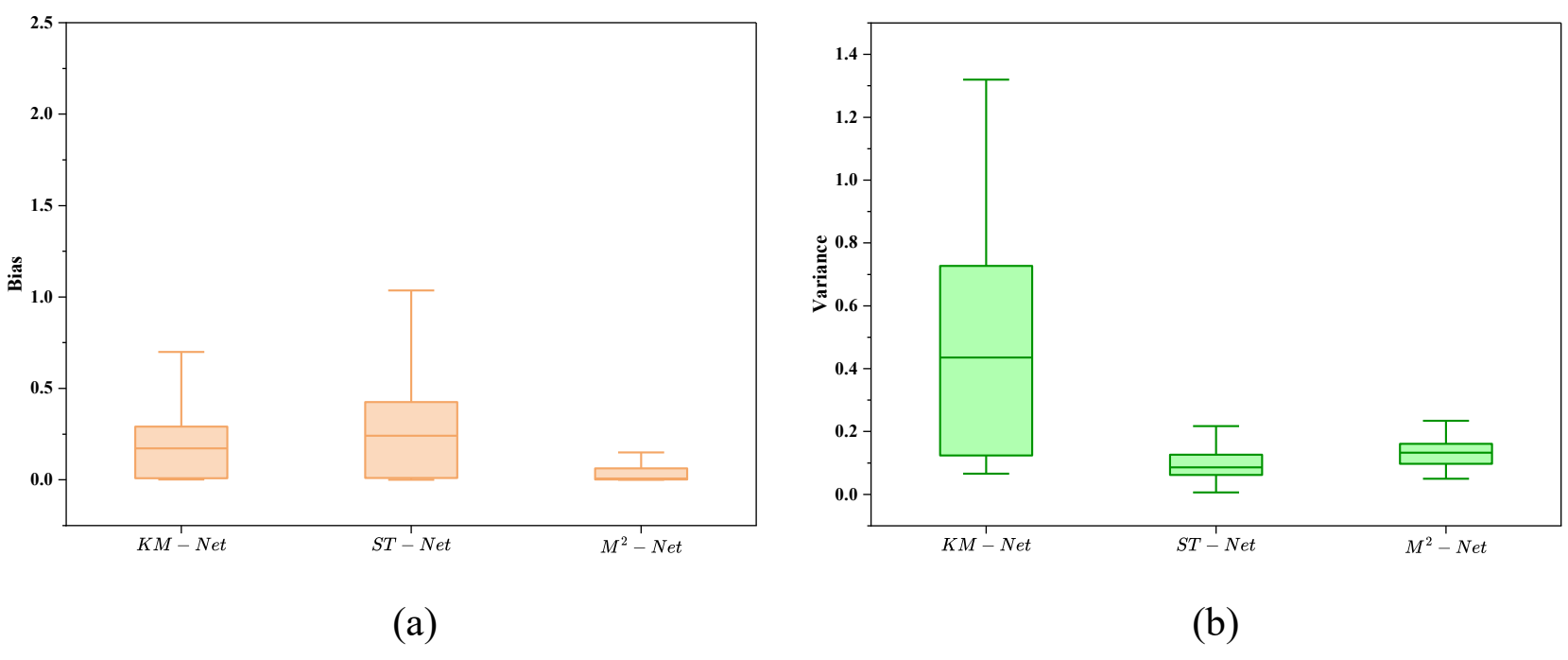

**Fig. 12.** Bias and variance box plots of $^{18}$F-FDG tracer in the subsequent frames predicted by ST-Net, KM-Net, and M²-Net on the real PET data (a) bias box plot and (b) variance box plot.

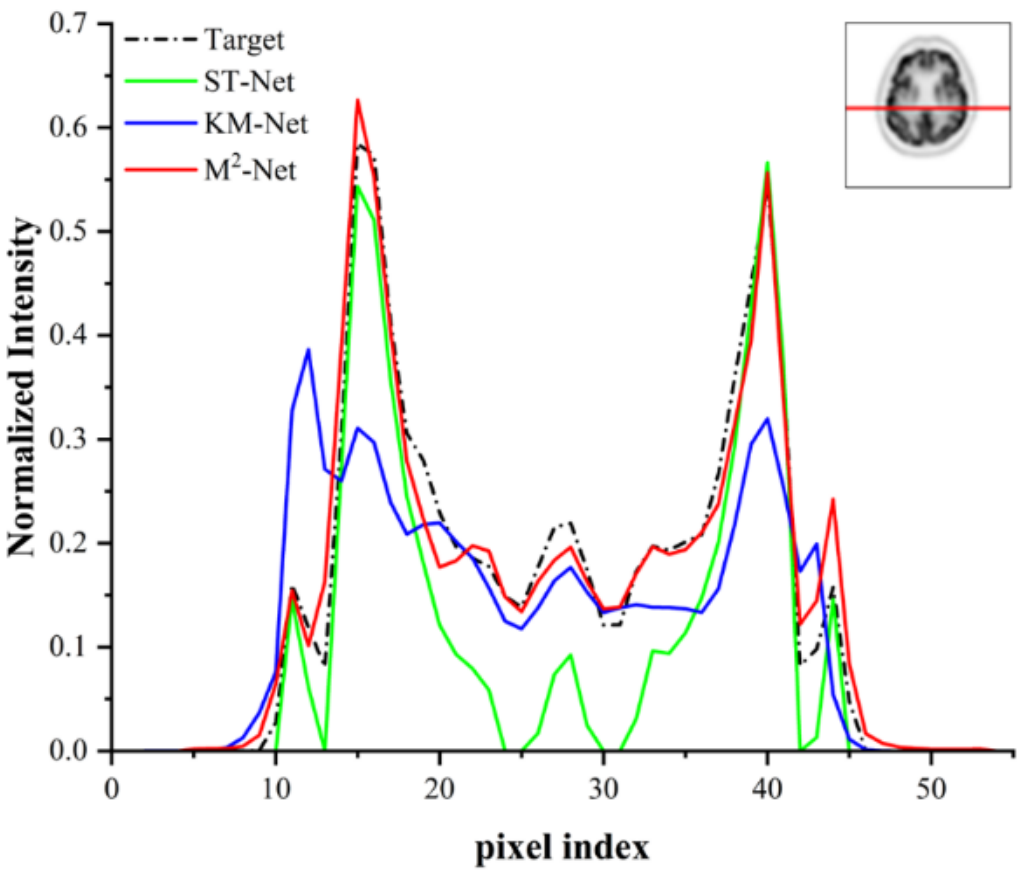

**Fig. 13.** Profile comparison results of three methods under the 17th frame prediction results of Clinical data.

*4.4. Ablation Study*

In this section, we performed ablation experiments on the loss function components of the proposed network. We used $^{11}$C-FMZ as the tracer to generate simulated data, with experimental data settings as described in Task 2. The results of the ablation experiments are shown in Table 5.

**Table 5**

Quantitative results of ablation experiments, Including MSE, PSNR, and SSIM. Lower MSE, higher SSIM, and greater PSNR are highlighted in bold.

| $\mathcal{L}_1$ | $\mathcal{L}_2$ | $\mathcal{L}_3$ | $\mathcal{L}_4$ | PSNR | SSIM | MSE |
|---|---|---|---|---|---|---|
| ✓ | – | – | – | 25.73 | 0.8016 | 0.0036 |
| ✓ | ✓ | – | – | 37.23 | 0.9837 | 0.0003 |
| ✓ | ✓ | – | ✓ | 38.73 | 0.9867 | 0.0002 |
| ✓ | ✓ | ✓ | ✓ | **42.67** | **0.9935** | **0.0001** |
| ✓ | – | ✓ | ✓ | 27.87 | 0.8471 | 0.0024 |
| – | ✓ | ✓ | ✓ | 20.67 | 0.6018 | 0.0106 |

Each row of the table indicates which loss components were included in the loss term, along with the corresponding PSNR, SSIM, and MSE metrics. From the second and third rows of the table, it can be seen that adding the $\mathcal{L}_4$. to the network yields better test metrics compared to using only $\mathcal{L}_1$ and $\mathcal{L}_2$. From the third and fourth rows, it can be observed that adding the $\mathcal{L}_3$ to the network results in better test metrics compared to using only $\mathcal{L}_1$, $\mathcal{L}_2$, and $\mathcal{L}_4$. These findings indicate that the losses $\mathcal{L}_3$ and $\mathcal{L}_4$, established from the final output, have a positive impact on network training. From the first, second and fifth rows, it is clear that removing the intermediate state output loss $\mathcal{L}_2$ significantly reduces the test metrics. Similarly, from the fourth, and sixth columns, it can be observed that for intermediate state output loss $\mathcal{L}_2$, the impact of missing $\mathcal{L}_1$ on the results is also significant This demonstrates the critical importance of the intermediate state output for the proposed $\text{M}^2$-Net . Both the intermediate state output loss $\mathcal{L}_1$ and $\mathcal{L}_2$ and the final output losses $\mathcal{L}_3$ and $\mathcal{L}_4$ positively contribute to enhancing the network's predictive performance.

To balance the coefficients of different loss terms, experiments were designed. Due to the extensive nature of the ablation experiments and the relatively minor impact of $\lambda$ on the results, we randomly selected 20% of the original dataset for this part of the experiments to accelerate the process. Since the $\mathcal{L}_4$ loss has a relatively small impact on improving network performance, this experiment fixes $\lambda_3$ at 1 and mainly explores the effect of different values of $\lambda_1$ and $\lambda_2$ on network performance. The experimental results are shown in Table 6. Based on the table, the final settings for $\lambda_1$ , $\lambda_2$ and $\lambda_3$ are 1.2, 1, and 1, respectively.

**Table 6**

Quantitative results of ablation experiments, Including MSE, PSNR, and SSIM. Lower MSE, higher SSIM, and greater PSNR are highlighted in bold.

| $\lambda_1$ | $\lambda_2$ | PSNR | SSIM | MSE |
|---|---|---|---|---|
| 1 | 1 | 40.77 | 0.9826 | **0.0001** |
| 1.2 | 1 | **41.91** | **0.9862** | **0.0001** |
| 1.4 | 1 | 41.53 | 0.9839 | **0.0001** |
| 1 | 1.2 | 41.17 | 0.9846 | **0.0001** |
| 1.2 | 1.2 | 41.37 | 0.9854 | **0.0001** |
| 1.4 | 1.2 | 39.24 | 0.9805 | 0.0002 |
| 1 | 1.4 | 41.47 | 0.9852 | **0.0001** |
| 1.2 | 1.4 | 39.56 | 0.9806 | 0.0002 |
| 1.4 | 1.4 | 39.53 | 0.9811 | **0.0001** |

## 5. Discussion

In clinical practice, obtaining kinetic parameters of compartment models is very important. Generating kinetic parameter images from dynamic PET images remains a challenge. Additionally, the long scan times of dynamic PET pose discomfort for patients and healthcare personnel. This paper introduces a method: using a multi-module network to generate kinetic parameters from the earlier frames of dynamic PET images, and then generating the complete dynamic PET images, significantly reducing the scan time.

In this study, the performance of $M^2$-Net, ST-Net, and KM-Net was compared. The robustness of $M^2$-Net , ST-Net, and KM-Net was evaluated through experiments using simulated data with different tracer types, different phantom types and different levels of noise. To make simulated data more realistic, we used structurally complex brain phantoms. To simulate individual variations, many sets of kinetic parameters were used in the generation process of simulated data. Based on the metrics and image presentations of the experimental results, it can be observed that in various scenarios, $M^2$-Net generated higher-quality dynamic PET images. In addition to simulation experiments, this study was also tested on clinical human brain data. $M^2$-Net continued to perform well in quantitative results, further demonstrating the proposed method's strong generalization capability. Compared to ST-Net and KM-Net, the advantages of $M^2$-Net can be attributed to the following points: This method combines deep learning techniques with traditional theories. Deep learning methods are used to construct learnable reversible modules, while traditional theories are used to construct non-learnable irreversible modules, including kinetic model algorithms and parametric image generation algorithms. The reversible module, as the core module of the network, employs invertible networks to achieve precise mapping from dynamic PET images to kinetic parameter images. Kinetic parameter images serve as intermediate state outputs, which not only provide a loss term based on themselves but also generate dynamic PET images and parametric images to add additional loss terms to the network. Multiple losses add constraints in various modalities, enabling the network to learn a more accurate mapping of kinetic parameter images.

However, there are some limitations in the current study: Currently, the effectiveness of our method may be constrained by the scarcity of training data. We anticipate that as the dataset size grows in the future, the model will become more adept at generalizing to other unseen datasets and a wider variety of tracers.

## 6. Conclusion

This work introduced the $M^2$-Net, a multi-module network that integrates deep learning methods with traditional theoretical algorithms. In this network, an end-to-end reversible network is used to construct a reversible module, and an algorithm based on traditional theory is used to construct an irreversible module. The two modules together add constraints to the optimization of the network. This network can accurately predict kinetic parameter images from the earlier frames of dynamic PET images and generate complete dynamic PET images based on kinetic model algorithms. This method significantly reduced the scan time for dynamic PET images. In this study, the performance of different methods is first explored on simulated data sets. In studies based on simulation data sets, the generated results of this method exhibit excellent PSNR, SSIM, and MSE values, surpassing the performance of existing ST-Net and KM-Net. Additionally, $M^2$-Net demonstrated robustness to different tracers, phantom types and levels of noise. This study also conducted generalization experiments on real data, and the results preliminarily validated the feasibility of this method for application to actual dynamic PET data.

## 7. CRediT authorship contribution statement

**Jie Sun:** Conceptualization, Data Curation, Formal Analysis, Investigation, Software, Writing-Original Draft, Writing-Review & Editing. **Junyan Zhang:** Conceptualization, Formal Analysis, Investigation, Writing-Original Draft, Writing-Review & Editing. **Qian Xia:** Supervision, Resources, Investigation, Formal analysis, Data Curation. **Chuanfu Sun:** Software, Formal analysis, Visualization. **Yumei Chen:** Investigation, Validation. **Yunjie Yang:** Supervision, Validation. **Huafeng Liu:** Supervision, Validation. **Wentao Zhu:** Conceptualization, Supervision, Funding acquisition, Project administration, Resources, Writing-Original Draft, Methodology, Writing-Review & Editing. **Qiegen Liu:** Conceptualization, Funding acquisition, Writing-Original Draft, Writing-Review & Editing, Project administration, Methodology, Supervision.

## 8. Declaration of competing interest

The authors declare that they have no known competing financial interests or personal relationships that could have appeared to influence the work reported in this paper.

## 9. Acknowledgments

This work was supported by National Natural Science Foundation of China (62001425, 62122033, 62571476) and in part by Key Research and Development Program of Zhejiang Province (2021C03029).